
\magnification\magstep1
\scrollmode

\headline={\ifnum\pageno=1\firstheadline\else
\ifodd\pageno\rightheadline \else\leftheadline\fi\fi}
\def\firstheadline{\hfil}
\def\rightheadline{\hfil}
\def\leftheadline{\hfil}
      \footline={\ifnum\pageno=1\firstfootline\else\otherfootline\fi}
\def\firstfootline{\rm\hss\folio\hss}
\def\otherfootline{\hfil}

\font\weerm=cmr10  scaled 833

\parindent=1.5truepc
\hsize=6.0truein
\vsize=8.5truein
\nopagenumbers


\font\tenbbb=msbm10 \font\sevenbbb=msbm7  
\newfam\bbbfam
\textfont\bbbfam=\tenbbb \scriptfont\bbbfam=\sevenbbb
\def\Bbb{\fam\bbbfam}      

\def\opname#1{\mathop{\rm#1}\nolimits} 

\def\a{\alpha}                    
\def\Al{{\cal A}}                 
\def\b{\beta}                     
\def\barox{\mathrel{\overline\otimes}} 
\def\Bl{{\cal B}}                 
\def\C{{\Bbb C}}                  
\def\d{{\bf d}}                   
\def\del{\partial}                
\def\delslash{{\partial\mkern-9mu/}} 
\def\E{{\cal E}}                  
\def\ga{\gamma}                   
\def\Gam{{\mit\Gamma}}            
\def\H{{\cal H}}                  
\def\HH{{\Bbb H}}                 
\def\L{{\cal L}}                  
\def\la{\lambda}                  
\def\Om{\Omega}                   
\def\ox{\otimes}                  
\def\R{{\Bbb R}}                  
\def\sepword#1{\qquad\hbox{#1}\quad} 
\def\stroke{\mathbin\vert}        
\def\tfrac#1#2{{\textstyle{#1\over#2}}} 
\def\thalf{\tfrac12}              
\def\Tr{\opname{Tr}}              
\def\tr{\opname{tr}}              
\def\Trs{\Tr^+}                   
\def\w{\wedge}                    
\def\x{\times}                    
\def\7{\dagger}                   
\def\8{\bullet}                   
\def\.{\cdot}                     
\def\:{\colon}                    
\def\<#1,#2>{\langle#1\stroke#2\rangle} 



\centerline{\bf ON MARSHAK'S AND CONNES' VIEWS OF CHIRALITY}
\vglue 6truemm

\centerline{\rm JOSE M. GRACIA-BONDIA}
\baselineskip=15truept
\centerline{\it Departamento de F\'{\i}sica Te\'orica, Universidad
                Aut\'onoma, Madrid 28049, Spain}
\baselineskip=12truept
\centerline{and}
\centerline{\it Departamento de F\'{\i}sica Te\'orica, Universidad
                de Zaragoza, Zaragoza 50009, Spain}
\vglue 5truemm

\centerline{\weerm ABSTRACT}
\vglue 2truemm
{\rightskip=3truepc
 \leftskip=3truepc
 \weerm\baselineskip=12truept\noindent
I render the substance of the discussions I had with
Robert E. Marshak shortly before his death, wherein the kinship between
the ``neutrino paradigm'' ---espoused by Marshak--- and the central
notion of K-cycle in noncommutative geometry (NCG) was found. In that
context, we give a brief account of the Connes--Lott reconstruction of
the Standard Model (SM).
\vglue 3truemm}

\vfil

\baselineskip=14truept

\leftline{\bf   1. Bob's last adventure}
\vglue 3truemm

I met Bob Marshak at a Texan barbecue. It was
mid-September of 1991. There was a workshop to celebrate the sixtieth
anniversary of his closest disciple, E. C. G. Sudarshan of the Center
for Particle Physics of the University of Texas at Austin. As the
physicists formed in line for the barbecue, I happened to
fill the place just before Bob. I turned to congratulate him on his
moving speech of the previous evening. Minutes later we were fast and
deep in conversation (whenever Bob heartily attacked the ``good stuff''
at the barbecue). It was a friendship, of sorts, {\it \`a premi\`ere
vue}.

Although an indefatigable traveller, Bob had never visited Spain. We
started together thinking about a trip that would allow him to sample
the Spanish cultural diversity and to meet some of the Spanish particle
physicists. We agreed that October 1992 would be a good time for
travelling to Spain. He would be going first to the World Fair in
Seville, spending a week in Andaluc{\'\i}a. Afterwards he was to
make a tour of several Spanish universities. On November 27 of 1991 he
wrote me from Virginia: ``My wife and I think that this Spanish trip
will compare in excitement with our first trip long ago (in 1953)
arranged by Amaldi for visits to the excellent Italian universities and
cultural treasures''.

Bob and Ruth Marshak indeed flew into Madrid on October the first of
1992. Then they departed for Granada and Seville. I met them in Madrid
on the 8th, at the station, upon the arrival of the bullet train from
Seville. I could not avoid noticing that Bob was in worse physical
condition that he had seemed to be in Texas one year before.
Nevertheless, he kept in reasonably good health and high spirits during
the trip. I believe he enjoyed it immensely. The old Spanish and Flemish
masters, Mir\'o and Picasso gave special pleasure to him and Ruth. I
witnessed his childlike gaiety and was enchanted by his love of life and
physics. Bob lectured on the triumphs of the SM of particle interactions
and gave us his personal recollections of the startup period in Particle
Physics. He was never far from a fax machine, meanwhile, as he was
giving the finishing touches to {\it Conceptual Foundations of Modern
Particle Physics},$^1$ fated to be his posthumous book.

\vskip 6truemm plus .5mm

\leftline{\bf   2. Bob's theoretical concerns at the end of his life}
\vglue 4truemm

During two unforgettable weeks in October 1992, Bob showered on me his
intimate knowledge of all theoretical aspects of particle physics. We
talked over breakfast, we talked on the trains, we talked over late
Spanish dinners. Some of the things he tried to explain to me I
understood only when I got {\it Conceptual Foundations\dots} in my
hands.$^1$ Others I will never fully grasp. Marshak's book is indeed a
superb conceptual legacy. All the challenging problems
associated with the SM are expounded with penetrating detail
and grouped in a coherent whole.

To report that Robert's scientific interests in the last period of his
life turned around the themes of his book will surprise no one. However,
there were favourites. He stressed how the original Marshak--Sudarshan
version of V--A invariance (in contrast to the Feynman--Gell-Mann one)
was based on the principle of chirality invariance, and he tried to
impress on me the importance of chirality and chiral gauge anomaly-free
constraints in modern particle dynamics.$^2$ He explained to me at
length the origins of the $U(1)$ and the ``strong CP" problems in QCD
and his solution$^3$ (proposed together with S. Okubo) to the latter.
He was eloquent on the advantages of the grand unification model based
on $SO(10)$. I cherish a very lucid account by Bob's own hand of
the earlier lepton-baryon symmetry, leading to the concepts of weak
hypercharge and weak isospin.

Bob also talked to me about his work as deputy leader of one of the
``theoretical'' groups during the atom bomb project at Los Alamos.
Robert and Ruth shared with me vivid memories about Klaus Fuchs, who
passed to the Soviets the secrets of the bomb. I came to a measure of
understanding and respect for the ethical convictions that led him to
assume the perilous presidency of New York City College. And this is how
I came into the privilege of being almost the last person to learn from
Bob.

In exchange, Bob asked me to report to him on the reconstruction of the
SM in the non-commutative geometry approach pioneered by
Connes and Lott. He was fascinated by NCG. During those lively
discussions, we realized that the ``neutrino paradigm''$^1$
that pervades Marshak's view of the SM and Connes' key
concept of $K$-cycle are like two sides of the same coin.

On the 25th of October I wished Robert and Ruth good travel on their
departure from Spain. Bob was contented and in an expansive mood. Some
time later I got a last letter from Bob. Little did I suspect
that we would not meet again by shade or sunlight.

\vskip 6truemm plus .5mm

\leftline{\bf   3. Chirality invariance and Noncommutative geometry}
\vglue 4truemm

I can do no better to pay homage to Bob than to deliver the substance of
the conversations we had on Connes' generalized geometry and the
Standard Model. Marshak contends that the chiral invariance of the Weyl
fermions plays a key r\^ole in the SM. Because of the large scale of the
spontaneous symmetry breaking mechanism that gives masses to the
fermions, it is expected that the departures from the ```neutrino
paradigm'' are small, except perhaps for the top quark. I will introduce
noncommutative geometry by considering a seemingly unrelated question:
the possibility of deriving the motion of a classical particle on a
manifold from the motion of quantum particles.

On a Riemannian manifold free particles move along geodesics.$^4$ A few
years ago, Connes realized that the simplest way to obtain geodesic
motion from quantum motion was to use neutrinos.

Connes' argument goes as follows.$^5$ Let ${\cal H} := L^2(S)$ be the
space of square integrable sections of the irreducible spinor bundle $S$
over the compact spin manifold~$M$, and $D$ the corresponding Dirac
operator. Recall that the algebra ${\cal A} = C^\infty(M)$ of smooth
(complex) functions over the manifold acts on ${\cal H}$ by
multiplication operators, i.e., multiplication by scalars on each fibre
of~$S$. The densely defined operator $[D,f]$, for $f \in {\cal A}$, is
bounded. Indeed, we have immediately $D(fs) - f\,Ds = c({\bf d}f)s$,
where $c({\bf d}f)$ means Clifford multiplication of the spinor~$s$
by~${\bf d}f$ and $\bf d$ denotes the ordinary differential of~$f$. This
operator is majorized by the supremum norm of~${\bf d}f$, which equals
the Lipschitz norm of~$f$, i.e., $\|f\|_{\rm Lip} := \sup_{p\neq q}
|f(p) - f(q)|/d(p,q)$, with $d(p,q)$ denoting the geodesic distance. The
geodesic distance is defined conventionally as the minimum path length
from $p$ to~$q$, but we can now turn the procedure around and recover
the metric on~$M$ from the Dirac operator and the algebra of functions
directly:
$$
d(p,q) :=
 \sup\ \{\, |f(p) - f(q)| : f \in {\cal A},\ \|[D,f]\| \leq 1 \,\}.
\eqno (1)
$$

Is it possible to derive the classical action from the kinematics of
quantum scalar particles? Indeed it should be, as the Laplacian encodes
the Riemannian geometry of the manifold. The formula is:
$$
d(p,q) :=
 \sup\ \{\, |f(p) - f(q)| : f \in {\cal A},\
            \thalf(\Delta f^2 + f^2 \Delta) -f \Delta f \leq 1 \,\}.
$$
The previous formula is given by Fr\"ohlich and Gaw\c edzki,$^6$ who
credit it to J. Derezi\'nski. The proof is the same, once one realizes
that left hand side of the inequality is the multiplication operator by
$\|{\bf d}f\|^2$. However, this is considerably more complicated.

Next, one can formalize the above into the key concept for
integrodifferential calculus in noncommutative geometry. By definition,
a $K$-cycle $({\cal H},D)$ on the $*$-algebra $\cal A$ consists
of a unitary representation of $\cal A$ on a Hilbert space $\cal H$,
together with an (unbounded) selfadjoint operator $D$ on~$\cal H$ with
compact resolvent, such that $[D,a]$ is bounded for all $a\in {\cal A}$.
We also assume $\cal H$ is a ${\bf Z}_2$-graded Hilbert space, equipped
with a grading operator $\Gamma$ such that $\Gamma^2 = 1$, that $\cal A$
acts on $\cal H$ by even operators, and that $D$ is an odd operator
(i.e., $a\Gamma = \Gamma a$ for $a \in {\cal A}$, and
$D\Gamma = - \Gamma D$). Then the right hand side of equation~(1) defines
also a distance on the space of states of the algebra (equipped with a
$K$-cycle), so it admits a natural noncommutative generalization.

\vfill\eject   

There is much more to it, from the physical point of view.
Noncommutative geometry comes into its own when we consider $K$-cycles
associated to finite algebras. Nothing more natural in the sequel
than to take up those finite algebras that give rise to the gauge
groups of particle physics and ``Dirac operators'' relating
the left- and right-handed representations of these algebras, just
like the standard Dirac operator relates the left- and right-handed
spinor representations. The matrix elements of these Dirac operators are
given by the Yukawa couplings among the fermions. It is possible
to combine both constructions to yield a Dirac--Yukawa operator, that
contains (in NCG) all the relevant information pertaining to the SM.
There is no way to figure out the mentioned parameters a priori.
Nevertheless, in contrast to the conventional version, the Higgs sector
(thus the boson mass matrix) is at the output end of Connes' machine
and the properties of the symmetry-breaking sector are entirely
determined. Indeed, the existence of the Higgs sector is a {\it
consequence\/} of chirality: it is the gauge field associated to the
intrinsic ``discreteness'' of the space that results from the existence
of left- and right-handed representations. This helps
to explain some characteristics of the Higgs field that are analogous
to those of nonabelian Yang--Mills fields. In particular, in that
reconstruction of the SM, the masses of the intermediate vector bosons
and of the Higgs particle are calculated, at least at the tree level,
in terms of the Yukawa couplings. They must be of the same order of the
top quark mass. We give some more details of the Connes--Lott setup in
the next two Sections.

\vskip 6truemm plus .5mm

\leftline{\bf   4. Connes' mathematical machine}
\vglue 4truemm

A ``noncommutative space'' is just a noncommutative algebra ${\cal A}$
(of operators on a Hilbert space). To get differential calculus on such
a space, one embeds $\cal A$ in a universal graded differential algebra
$\Omega^\bullet{\cal A} = \bigoplus_{n\geq0} \Omega^n{\cal A}$ generated
by symbols $a_0\,da_1\dots da_n$ with a derivation $d$ satisfying
$d(a_0\,da_1\dots da_n) = da_0\,da_1\dots da_n$, $d(1) = 0$,
$d^2 = 0$. This is an $\cal A$-bimodule: we multiply
$a_0\,da_1\dots da_n$ by $b \in {\cal A}$ on the right by applying the
rule $(da)b = d(ab) - a\,db$ repeatedly.

For the commutative case ${\cal A} = C^\infty(M)$, the smooth sections
of a hermitian vector bundle on~$M$ form a (right) module $\cal E$ over
the algebra~$\cal A$, which is of the form $p{\cal A}^m$ with
$p^2 = p = p^*$ in some $m \times m$ matrix algebra over~$\cal A$;
moreover, $\cal E$ carries a positive hermitian form $(\cdot,\cdot)$ with
values in~$\cal A$. Such modules, over more general algebras, are
``noncommutative vector bundles''. A compatible {\it connection\/} on
$\cal E$ is then a linear map
$\nabla \colon {\cal E} \to {\cal E} \otimes_{\cal A} \Omega^1{\cal A}$
satisfying $\nabla(sa) = (\nabla s) a + s \otimes da$ and
$d(s,s') = (\nabla s, s') + (s,\nabla s')$, for $s,s' \in {\cal E}$,
$a \in {\cal A}$. Its curvature is the matrix-valued 2-form $\theta$
given by $\nabla^2(s) = \theta s$. Gauge transformations
$\nabla \mapsto u \nabla u^*$ are given by unitary matrices $u$ over
$\cal A$ satisfying $up = pu$; thus the utility of the vector bundle
$\cal E$ is to specify the gauge group.

Integration over a noncommutative space is given by the ``Dixmier
trace'' of compact operators on a Hilbert space~$\cal H$. A positive
compact operator $A$ with eigenvalues
$\lambda_1 \geq \lambda_2 \geq \dots\,$ lies in the Dixmier trace class
if and only if $\lambda_1 + \cdots + \lambda_n = O(\log n)$. The
Dixmier trace is a generalized limit of the form
$$
\mathop{\rm Tr}\nolimits^+ A
 := \lim_{n\to\infty} {\lambda_1+\cdots+\lambda_n \over \log n},
$$
which can be extended to a linear functional on the full Dixmier trace
class; we have $\mathop{\rm Tr}\nolimits^+ T = 0$ for $T$ in the
ordinary trace class.

When ${\cal H} = L^2(S)$ is a spinor bundle over a compact
Riemannian manifold $M$ of even dimension $n = 2m$, and
$D = \gamma^\mu \partial_\mu$ is the Dirac operator, then $|D|^{-n}$
lies in the Dixmier trace class, and a fundamental trace theorem of
Connes$^{7,8}$ yields the following integral formula, for
$a \in C^\infty(M)$:
$$
\mathop{\rm Tr}\nolimits^+ (a |D|^{-n})
 = {1\over m!\,(2\pi)^m} \int_M a(x) \,d\mathop{\rm vol}(x).
\eqno (2)
$$
This is how the Dixmier trace, in the presence of a $K$-cycle, gives a
precise generalization of integration over a manifold.

The $K$-cycle also allows us to refine the ``differential calculus'' by
reducing the large differential algebra $\Omega^\bullet{\cal A}$ to a
more useful one. We can represent $\Omega^\bullet{\cal A}$ on~$\cal H$
by taking
$$
\pi(a_0\,da_1\dots da_n) := i^n\, a_0\,[D,a_1] \dots [D,a_n].
$$
One can have $\pi(b) = 0$ with $\pi(db) \neq 0$. We must factor out the
differential ideal of ``junk'' $J :=
 \{\, b'+db'' \in\Omega^\bullet{\cal A} : \pi b' = \pi b'' = 0\,\}$,
thereby obtaining a new graded differential algebra of ``$D$-forms'' by
$$
\Omega_D^\bullet {\cal A}
 := \pi(\Omega^\bullet{\cal A})/J \equiv \pi_D(\Omega^\bullet{\cal A}).
$$
For the Dirac $K$-cycle, the quotient algebra
$\Omega_D^\bullet C^\infty(M)$ is the usual algebra of
differential forms on~$M$.

``Universal'' connections and curvatures on $\Omega^\bullet{\cal A}$
pass to connections and curvatures on $\Omega_D^\bullet{\cal A}$. We can
integrate the square of the curvature $\theta$ to get a gauge-invariant
and nonnegative functional
$I(\nabla) := \mathop{\rm Tr}\nolimits^+ (\pi(\theta)^2\,|D|^{-n})$ on
(universal) connections. Factoring out the unwanted junk~$J$ is
accomplished by a certain orthogonal projection $P$; if we start from a
connection $\widetilde\nabla$ defined with $D$-forms, we can set
$$
YM(\widetilde\nabla) := \|P\pi(\theta)\|^2
 = \inf \{\, I(\nabla) : \pi_D(\nabla) = \widetilde\nabla \,\}.
\eqno (3)
$$
In the commutative Riemannian case, $\widetilde\nabla$ is given by an
ordinary 1-form $\omega$ on~$M$, and the trace theorem (2) gives
$$
\|P\pi(\theta)\|^2 = {(2\pi)^{-n/2}\over (n/2)!}
 \int_M \|{\bf d}\omega\|^2 \,d\mathop{\rm vol},
$$
which is the classical Yang--Mills action.

\vskip 6truemm plus .5mm

\leftline{\bf   5. Reconstructing the Standard Model}
\vglue 4truemm

We take, as algebras and Hilbert space for the model:
$$
\eqalign{
\Al &:= C^\infty(M,\R) \ox_\R (\C \oplus \HH)
 \cong  C^\infty(M,\C) \oplus C^\infty(M,\HH);
\cr
\Bl &:= C^\infty(M,\R) \ox_\R (\C \oplus M_3(\C))
 \cong  C^\infty(M,\C) \oplus M_3(C^\infty(M,\C));
\cr
\H  &:= L^2(S) \ox \H_F,
\cr}
$$
where $\H_F$ is a finite dimensional Hilbert space carrying commuting
representations of the ``finite-part'' algebras
$$
\Al_F := \C \oplus \HH,  \qquad  \Bl_F := \C \oplus M_3(\C).
$$

The representation $\pi$ of $\Al$ on~$\H$ decomposes into
representations $\pi_\ell \oplus \pi_q$ on the lepton and quark sectors:
$\H = \H_\ell \oplus \H_q$. Likewise, the representation $\sigma$ of
$\Bl$ on~$\H$ decomposes into $\sigma_\ell \oplus \sigma_q$. We take
$\sigma_\ell(\mu,B) := \mu I$ on $\H_\ell$, for $(\mu,B) \in \Bl$ (no
colouring of leptons), and $\sigma_q(\mu,B) = \sigma'(B)$, where
$\sigma'$ is a {\it faithful\/} representation of $M_3(C^\infty(M,\C))$.
Thus $\H_q$ splits as $\H_q = \H_1 \oplus \H_1 \oplus \H_1
 = \H_1 \ox \C^3$. Since $\pi_q(\Al)$ must commute with $\sigma'(\Bl)$,
we have $\pi_q = \pi_1 \ox I_3$ where $\pi_1$ is a representation of
$\Al$ on $\H_1$. Writing $\pi_0 = \pi_\ell$, we arrive at
$$
\pi(\la,q) = \pi_0(\la,q) \oplus
             \pi_1(\la,q) \oplus \pi_1(\la,q) \oplus \pi_1(\la,q),
 \sepword{for} (\la,q) \in \Al.
$$
Here $\pi_0$, $\pi_1$ are independent real representations of $\Al$.

$\H$ can be {\it graded\/} so that both $\Al$ and $\Bl$ act by even
operators. The grading operator is $\Gam := \pi(1,-1)$, so $\pi(\la,q)$
has a block matrix form over $\H = \H_R \oplus \H_L$. We take
$$
\pi_0(\la,q) :=
 \pmatrix{\la & 0 & 0\cr 0 &\a & \b\cr 0 &-\b^* & \a^*\cr} \ox I_{N_G},
\quad
\pi_1(\la,q) :=
 \pmatrix{\la & 0 & 0 & 0 \cr 0 & \bar\la & 0 & 0 \cr
          0 & 0 & \a & \b \cr 0 & 0 & -\b^* & \a^*\cr} \ox I_{N_G},
$$
where $N_G$ is the number of particle generations.

The operator $D$ which gives the $K$-cycle must act independently on
each of~$\H_\ell$ and~$\H_q$; otherwise, the matrix $[D, \pi(\la,q)]$
will contain cross-terms not commuting with all $\sigma(\mu,B)$. This
condition forces $D$ to be of the form
$D = D_0 \oplus D_1 \oplus D_1 \oplus D_1$, where $D_0$, $D_1$ are odd
operators on $\H_0$, $\H_1$ respectively.

   If we apply this scheme to the ``finite-part'' algebras only, we
retrieve matrix operators $D_{F0}$ on~$\H_{F0}$ and $D_{F1}$
on~$\H_{F1}$, of the form
$$
D_{Fj} = \pmatrix{0 & G_j^\7 \cr G_j & 0 \cr}
$$
with respect to the right-left splitting, where $G_0$, $G_1$ are
suitable complex matrices. Specifically, we have
$$
G_1 = \pmatrix{g_d & 0 \cr 0 & g_u \cr},  \qquad
G_0 = \pmatrix{g_e \cr 0 \cr},  \qquad
$$
where $g_d, g_u, g_e \in M_{N_G}(\C)$.

We now take the {\it graded\/} tensor product of the $K$-cycles
$(C^\infty(M,\R), L^2(S), \delslash)$ and $(\Al_F, \H_F, D_F)$. The
$K$-cycle $(\Al, \H, D)$, with $\Al := C^\infty(M,\R) \ox \Al_F$,
$\H := L^2(S) \ox \H_F$, is given by:
$$
D := (\delslash \ox I) \oplus (1 \ox D_F),
$$
and we stipulate that the graded differential algebra $\Om_D(\Al)$ be
defined as the {\it graded\/} tensor product of algebras:
$$
\Om_D^\8(\Al)
 := \Om_\delslash^\8(C^\infty(M,\R)) \barox \Om_{D_F}^\8(\Al_F).
$$
This amounts to the rule that, for $f \in C^\infty(M,\C)$ and
$(\la,q) \in \Al_F$, $c(\d f) = \ga^\mu \del_\mu f$
anticommutes with $\delta(\la,q) := [D_F, \pi_F(\la,q)]
 = \pmatrix{0 & G^\7(q - \la) \cr (\la - q)G & 0 \cr}$.

   We have
$\Om_D^0(\Al) \simeq \Al \simeq C^\infty(M,\C) \oplus C^\infty(M,\HH)$.
Next, $\Om_D^1(\Al)$ is generated by elements of the form
$(f_0 c(\d f_1), r_0 c(\d r_1)) + (f_2,r_2) \,\delta(\la_1,q_1)$, where
$f_j \in C^\infty(M,\C)$, $r_j \in C^\infty(M,\HH)$. Schematically, we
may write
$$
\Om_D^1(\Al) = \pmatrix{\E^1(M,\C) &  C^\infty(M,\HH) \cr
                        C^\infty(M,\HH) & \E^1(M,\HH) \cr},
$$
where $\E^k$ denotes (ordinary) $k$-forms. To determine
$\pi(\Om^2(\Al))$, we notice that
$$
\pmatrix{\d f_1 & G^\7 s_1 \cr r_1 G & \d q_1 \cr}
 \pmatrix{\d f_2 & G^\7 s_2 \cr r_2 G & \d q_2 \cr}
 = \pmatrix{\d f_1\cdot\d f_2 + G^\7 s_1r_2 G
         & G^\7 (s_1\,\d q_2 - \d f_1\,s_2)  \cr
           (\d q_1\,r_2 - r_1\,\d f_2) G
         & \d q_1\cdot\d q_2 + r_1 GG^\7 s_2 \cr},
$$
with the dot denoting Clifford multiplication; the anticommutation rule
enables the $\d f_j$ to slip past the matrices $G$ or $G^\7$ with a
change of sign.

   To find $\Om_D^2(\Al)$, we must identify and factor out the junk
subspace $J^2$. Two independent scalar terms in this subspace drop from
$\d f_1\cdot\d f_2$ and $\d q_1\cdot\d q_2$; another term arises from
the relation
$$
GG^\7 \pmatrix{\a & \b \cr -\b^* & \a^* \cr}
 = \pmatrix{\a & \b \cr -\b^* & \a^* \cr} \ox (GG^\7)_+
   + \pmatrix{\a & \b \cr \b^* & -\a^* \cr} \ox (GG^\7)_-,
$$
with $(G_0G_0^\7)_\pm = \thalf(g_eg_e^\7)$,
$(G_1G_1^\7)_\pm = \thalf(g_dg_d^\7 \pm g_ug_u^\7)$. The
``antiquaternionic'' second term on the right lives in $J^2$. A full
computation shows that the elements of $J^2$ are
$$
\pmatrix{\psi \ox I & 0 \cr
         0 & \chi \ox I + \tau \ox (GG^\7)_- \cr}
$$
where $\psi$, $\chi$, $\tau$ are respectively complex,
quaternionic and antiquaternionic-valued functions on~$M$. We can
identify $\Om_D^2(\Al)$ with the orthogonal complement of~$J^2$, i.e.,
we can ``subtract off'' the junk terms, and express an element of
$\pi_D(\Om^2(\Al))$ as
$$
\pmatrix{\d f_1 \w \d f_2 + (G^\7 s_1r_2 G)_\bot
         & G^\7 (s_1\,\d q_2 - \d f_1\,s_2)  \cr
           (\d q_1\,r_2 - r_1\,\d f_2) G
         & \d q_1 \w \d q_2 + (r_1 s_2 \ox (GG^\7)_+)_\bot \cr},
\eqno (4)
$$
where the subindex $\perp$ on a matrix indicates that its trace has been
subtracted out.

We may express $\Om_D^2(\Al)$ schematically as:
$$
\pmatrix{\E^2(M,\C) \oplus C^\infty(M,\HH) & \E^1(M,\HH) \cr
         \E^1(M,\HH) & \E^2(M,\HH) \oplus C^\infty(M,\HH) \cr},
$$
with the following multiplication rule for
$\Om_D^1(\Al) \x \Om_D^1(\Al) \to \Om_D^2(\Al)$:
$$
\pmatrix{A_1 & s_1 \cr r_1 & V_1 \cr}
\pmatrix{A_2 & s_2 \cr r_2 & V_2 \cr}
= \pmatrix{A_1 \w A_2 \oplus s_1r_2 &  s_1 V_2 - A_1 s_2 \cr
           V_1 r_2 - r_1 A_2        &  V_1 \w V_2 \oplus r_1 s_2 \cr}.
$$

The differentials $d\: \Om_D^0(\Al) \to \Om_D^1(\Al)$ and
$d\: \Om_D^1(\Al) \to \Om_D^2(\Al)$ are given by:
$$
d(f,q) := \pmatrix{\d f & q - f \cr  f - q & \d q \cr},  \quad
d\pmatrix{A & s \cr r & V \cr}
 = \pmatrix{\d A \oplus (r + s) & -\d s - A + V \cr
            \d r - A + V        & \d V \oplus (r + s) \cr}.
$$

Similar arguments apply to the algebra $\Bl$; we have
$\Om_D^0(\Bl) \simeq \Bl$ and $\Om_D^1(\Bl) \simeq \sigma(\Om^1(\Bl))$.
Since $1 \ox D_F$ commutes with $\sigma(\Bl)$, the only junk arises from
the scalar level in Clifford algebra; its removal yields:
$$
\Omega_D^2(\Bl) \simeq \E^2(M,\C) \oplus M_3(\E^2(M,\C)).
$$

We can add skewsymmetric 1-forms $\a \in \Om_D^1(\Al)$ and
$\b \in \Om_D^1(\Bl)$ by identifying these modules with their (faithful)
representations on~$\H$. Thus we consider $\a + \b$ as a connection
form. The total curvature is given by
$$
\theta = (d\a + \a^2) + (d\b + \b^2) = \theta_\a + \theta_\b,
$$
since the cross-terms cancel. Take
$$
\a = \pmatrix{A & r^* \cr r & V \cr}, \qquad
\b = \pmatrix{A' & 0  \cr 0 & K \cr},
$$
where $A,A',V,K$ are antisymmetric 1-forms with respective values in
$\C$, $\C$, $\HH$ and $M_3(\C)$.

   Reduction of the gauge group from
$U(1) \x SU(2) \x U(1) \x U(3)$ to $SU(2) \x U(1) \x SU(3)$ is effected
by the algebraic chirality condition:
$$
\Tr_{\H_L}(\a + \b) = 0,  \qquad  \Tr_{\H_R}(\a + \b) = 0.
$$
Now $V^* = -V$ means $V$ is a zero-trace quaternion, so
$\Tr_{\H_L}(\a) = 0$ automatically; thus $\Tr_{\H_L}(\b) = 0$, which
yields the condition $A' = - (K_{11} + K_{22} + K_{33})$. Moreover,
$$
\Tr_{\H_R}(\a + \b)
 = N_G(A + A') + 3N_G(A - A) + 2N_G(K_{11} + K_{22} + K_{33}),
$$
on separating the lepton and quark sectors; thus
$A + A' + 2(K_{11} + K_{22} + K_{33}) = 0$. Combining both conditions,
we get the chirality reduction rule:
$$
A = A' = -(K_{11} + K_{22} + K_{33}).
$$

\smallskip

The bosonic Yang--Mills functional (3) for the model may now be
computed, yielding the Lagrangian $\L$ from the trace theorem:
$I(\nabla) = \int_M \L$. The several components of $\theta$ contribute
various terms of the Lagrangian; two-forms on~$M$ yield the pure-gauge
part, one-forms yield the kinetic term for the Higgs field and its
coupling with the flavour gauge bosons, and functions on~$M$ give the
Higgs self-interaction term.

The appearance of the Higgs terms in the Lagrangian happens as follows.
We write $q = 1 + r$ where $r$ is the quaternionic function in
equation~(4). It turns out that $\theta_\a$ depends on~$q$ only through
the expressions ${\bf D}q = \d q - qA + Vq$ (a covariant derivative) and
$(qq^* - 1)$. We then interpret $q$ as a Higgs doublet:
$$
q = \pmatrix{\Phi_1 & - \Phi_2^* \cr \Phi_2 & \Phi_1^* \cr}
 = \Phi_1 - \Phi_2^* j.
$$
Then $\L$ is of the general form: pure-gauge part
${}+ C_1(D_\mu\Phi)(D^\mu\Phi) + C_0(\|\Phi_1\|^2 + \|\Phi_2\|^2 - 1)^2$.

Two aspects of this Lagrangian must be remarked. Firstly, the reduction
rules affect mainly the coefficients of the pure gauge terms. Secondly,
there is some freedom in selecting the exact form of the Dixmier trace
one must use. We can use
$\Trs = \a_\ell \Trs_\ell + \a_q \Trs_q$, with $\a_\ell + \a_q = 1$.
These coefficients enter the junk components $\psi$, $\chi$
of~$\theta_\a$, and thereby enter the Lagrangian in a nonlinear way. The
result of this computation has been given by Kastler and Sch\"ucker.$^9$
After identification with the usual notations for the gauge fields, it
is:
$$
\eqalign{
\L &= - N_G(3\a_\ell + \tfrac{11}3 \a_q) F_{\mu\nu} F^{\mu\nu}
      - N_G(\tfrac14 \a_\ell + \tfrac34 \a_q) H_{\mu\nu}^a H_a^{\mu\nu}
      - N_G \a_q G_{\mu\nu}^a G_a^{\mu\nu}
\cr
&\quad + 2(\a_\ell \tr(g_e^\7g_e) + 3\a_q \tr(g_d^\7g_d + g_u^\7g_u))
    (D_\mu\Phi)(D^\mu\Phi) + (\|\Phi_1\|^2 + \|\Phi_2\|^2 - 1)^2 \times
\cr
&\qquad \times \biggl[ \tfrac32 \a_\ell \tr(g_e^\7g_e)^2
    + \tfrac92 \a_q \tr((g_d^\7g_d)^2 + (g_u^\7g_u)^2)
    + 3\a_q \tr(g_d^\7g_dg_u^\7g_u)
\cr
&\qquad - {1\over N_G}
  \Bigl( {1\over \a_\ell + 6\a_q} + {1\over 2\a_\ell + 6\a_q} \Bigr)
  (\a_\ell \tr(g_e^\7g_e) + 3\a_q \tr(g_d^\7g_d + g_u^\7g_u))^2 \biggr].
\cr}
\eqno (5)
$$
There seems to be no reason, at present, to take $\a_\ell$ and $\a_q$
different from~$\thalf$.

\bigskip

   Therefore, NCG suggests values for the masses of undiscovered
particles. From the expression~(5), we obtain immediately
$m_W = \sqrt{C_1/4 N_G}$, from which the mass of the top quark is
estimated to be $m_t = 160.4$~GeV. Also, the mass of the Higgs particle
would be given by $m_H = 2\sqrt{C_0/C_1}$; one gets $m_H = 251.7$~GeV.

   It has been shown that parameter restrictions like the above (coming
from noncommutative geometry models) do not survive quantum
corrections.$^{10}$ On the other hand, if one adopts the point of view
that these restrictions are to be interpreted as tree-level constraints,
and as such are implemented in a mass-independent scheme at a given
energy scale, it is found that the physical predictions on the top and
Higgs masses depend fairly weakly on the aforementioned energy
scale.$^{11}$

\vglue 6truemm plus .5mm

\leftline{\bf Acknowledgment}

\vglue 4truemm

Many thanks to Joseph C. V\'arilly for several useful discussions, and
for some {\TeX}nical help.

\vglue 6truemm plus .5mm

\leftline{\bf   References}

\vglue 4truemm

\itemitem{1.} R. E. Marshak,
{\it Conceptual Foundations of Modern Particle Physics},
World Scientific, Singapore, 1993.

\itemitem{2.} C. Q. Geng and R. E. Marshak,
{\it Phys.\ Rev}. {\bf D39} (1989) 693.

\itemitem{3.} R. E. Marshak and S. Okubo,
{\it Prog.\ Theor.\ Phys}. {\bf 87} (1992) 1059.

\itemitem{4.} R. Abraham and J. E. Marsden,
{Foundations of Mechanics}, Benjamin, Reading, Massachusetts,
1978.

\itemitem{5.} A. Connes,
{\it Ergod.\ Thy.\ Dynam.\ Sys}. {\bf 9} (1989) 207.

\itemitem{6.} J. Fr\"ohlich and K. Gaw\c edzki,
lecture given at the Mathematical Quantum Theory conference held at the
University of British Columbia, Vancouver, August 1993.

\itemitem{7.} A. Connes,
{\it Commun.\ Math.\ Phys}. {\bf 117} (1988) 673.

\itemitem{8.} J. C. V\'arilly and J. M. Gracia-Bond{\'\i}a,
{\it J. Geom.\ Phys}. {\bf 12} (1993) 223.

\itemitem{9.} D. Kastler and T. Sch\"ucker, Theor.\ Math.\ Phys.\
{\bf 92} (1992) 522.

\itemitem{10.} E. Alvarez, J. M. Gracia-Bond{\'\i}a and C. P.
Mart{\'\i}n, {\it Phys.\ Lett}. {\bf B 306} (1993) 55.

\itemitem{11.} E. Alvarez, J. M. Gracia-Bond{\'\i}a and C. P.
Mart{\'\i}n, {\it Phys.\ Lett}.~{\bf B}, to appear.

\bye